\documentclass[10pt,conference,,twocolumn]{IEEEtran}
\usepackage[dvips]{graphicx}
\DeclareGraphicsExtensions{.eps}
\usepackage{amsmath}
\usepackage{amssymb}
\usepackage{amsthm}
\usepackage{xcolor}

\newcommand{\Expect}{{\rm I\kern-.5em E}}

\usepackage{mathtools}
\DeclarePairedDelimiter\ceil{\lceil}{\rceil}

\begin{document}

\title{Advanced Signal Processing Techniques for Fixed and Mobile Satellite Communications
}

\author{\IEEEauthorblockN{Pol Henarejos\IEEEauthorrefmark{1}, Ana P\'{e}rez Neira\IEEEauthorrefmark{1}, Nicol\`o Mazzali\IEEEauthorrefmark{2}, 
Carlos Mosquera\IEEEauthorrefmark{3}}
\IEEEauthorblockA{\IEEEauthorrefmark{1}Centre Tecnol\`{o}gic de Telecomunicacions de Catalunya\\
08860 Castelldefels, Barcelona, Spain\\
Email: \{pol.henarejos, ana.perez\}@cttc.es}
\IEEEauthorblockA{\IEEEauthorrefmark{2}Interdisciplinary Centre  for  Security,  Reliability  and  Trust  (SnT),\\
University  of  Luxembourg\\
Email: nicolo.mazzali@uni.lu}
\IEEEauthorblockA{\IEEEauthorrefmark{3}Signal Theory and Communications Department,\\
University  of  Vigo, 36310 Vigo, Spain\\
Email: mosquera@gts.uvigo.es\\
}}

\maketitle

\begin{abstract}
Enabling ultra fast systems has been widely investigated during recent decades. Although polarization is deployed from the beginning in satellite communications, nowadays it is being exploited for increasing the throughput of satellite links. More precisely, the irruption of multiple-input multiple-output (MIMO) technologies combined with polarization domain is a promising topic to provide reliable, robust and fast satellite communications. Better and more flexible spectrum use is also possible if transmission and reception can take place simultaneously in close or even overlapping frequency bands. In this paper we investigate novel signal processing techniques to increase the throughput of satellite communications in fixed and mobile scenarios. First, we analyse and investigate 4D constellations for the forward link. Second, we focus on the mobile scenario and introduce an adaptive algorithm which selects the optimal tuple of modulation order, coding rate and MIMO scheme that maximizes the throughput constraint to a maximum packet error rate. Finally, we describe   the operation of radio transceivers which cancel actively the self-interference posed by the transmit signal when operating in full-duplex mode.
\end{abstract}


\IEEEpeerreviewmaketitle

\section{Introduction}
In recent years, the increasing demand of higher data rate communications has motivated many researchers to focus their effort on the investigation of diversity techniques, such as massive multiple-input multiple-output (MIMO) systems \cite{Ma10} and space-time codes (STC) \cite{TaSeCa98}. However, both these techniques do not fit well the satellite scenario: the spatial richness required by MIMO systems cannot be provided by fixed satellite links \cite{ArLiBePaCoDe11}, and the delay caused by STCs is not acceptable in systems having time constraints. Nevertheless, polarization diversity has been considered as a viable option in S-band applications \cite{KyHuYlByShArGr14}, \cite{ShArOt12}. The signal processing described in these works is performed after the modulation, making it agnostic to the adopted constellation.

On the contrary, several results on constellation design can be found in the terrestrial MIMO literature, e.g., in \cite{WuLaPa06} and references therein. In most of these works, the adopted performance metric is the pairwise error probability and the union bound.

In the first part of the paper, we focus on assessing the performance of constellations for dual-polar (DP) satellite systems serving fixed users. In particular, we investigate four-dimensional (4D) constellations, where the number of dimensions is given by the total number of components (in-phase and quadrature) over the two polarizations. Moreover, a joint processing of the two streams is considered at the receiver as in a traditional dual-polar MIMO scheme. Unlike the cited works on constellation design in MIMO systems, we choose as main performance metrics the achievable information rate (AIR) and the pragmatic achievable information rate (PAIR). The PAIR allows for a joint evaluation of the performance of the constellation symbols and their labels, which is of paramount importance in practical scenarios (i.e., where channel coding is used). Indeed, most of the recent works on 4D constellations assess the performance in terms of symbol error rate or in uncoded systems \cite{YoWeYaBaKo13}, neglecting to take into account the impact of the labelling design.

In general, multidimensional constellations show better performance than 2D constellations \cite{LaFaTr94}. For example, QAM constellations over AWGN channel suffer from a loss with respect to the capacity of 1.53 dB. This shaping loss can be partially compensated by using a 4D constellation based on lattices. Indeed, the asymptotic shaping gain provided by a 4D lattice-based design is only 0.46 dB \cite{LaFaTr94}. The densest lattice in 4D is known, and its characteristics have been thoroughly studied in \cite{CoSl99}. Nevertheless, dense lattice-based constellations may perform poorly in practical systems because they maximize only the minimum Euclidean distance between symbols, which is a good design criterion only at high signal-to-noise ratios (SNRs).

In the following we compare the performance in terms of PAIR of 4D constellations over the AWGN channel. In particular, we will consider lattice-based constellations (also called lattice amplitude modulation, LAM), enhanced polypolarization modulation (EPPM, as described in \cite{YoWeYaBaKo13}), and constellations generated by means of the Cartesian product of two standard 2D constellations, denoted as as $\sqrt{M}\!\times\!\sqrt{M}$-QAM. The last approach is equivalent to transmitting independently over the two polarizations or performing spatial multiplexing \cite{ArLiBePaCoDe11}.

In the second part of the paper, we study the viability of DP in mobile scenarios. Although DP has been used for many decades in fixed satellite communications, the polarization multiplexing was performed without any adaptation nor flexibility and DP was not received simultaneously. 

However, it has been proven that DP can also be applied to mobile satellite communications. In this way, DP may be employed to increase the system capacity to increase the throughput of the individual links and increase the number of User Terminals (UE) connected to the network by taking the advantage of the partial decorrelation of the two polarizations. This approach is modelled using MIMO notation and it can be exploited by MIMO signal processing techniques.

The first challenge of DP is to provide a new communication system where the information can be modulated on the polarization state of the waveform and satisfy the scenario constraints. To achieve it, the terminals are able to adapt to the satellite channel and feedback to the ground gateway which modulation and coding scheme is the best for the session as well as which polarization MIMO scheme should be used. 

The second challenge is to implement the proposed algorithms in realistic scenarios. To fulfil it, we aim to deploy an adaptive algorithm and use the Broadband Global Area Network (BGAN) standard, specified in~\cite{ETSI}, as a benchmark. This standard describes procedures which provide multimedia mobile satellite communications with low latency and high flexibility in terms of throughput. Due to the long and slow shadowing, it is necessary to implement the physical layer abstraction (PLA) of the proposed scheme. Thus, the PLA is a tool to model the PHY, obtain these parameters that are involved in the adaptation of the link and estimate the error rate, without to run the whole coding and decoding chain.

Within the framework of the SatNEx consortium  special attention has been also put on the comparison  between full-duplex and half-duplex operation in the satellite for the operation of the new VHF Data Exchange System (VDES) standard \cite{IALA}. Full Duplex (FD) represents an attractive solution to improve the throughput of wireless communications. The term FD is historically used to refer to those systems that transmit and receive simultaneously, like Frequency-Division Duplexing (FDD). If transmission and reception take place in the same frequency band, then In-Band Full Duplex (IBFD) is the right term. We will present some initial considerations on the coexistence of simultaneous transmission and reception when leakage from the transmitter affects the receiver; some cancellation techniques are common also to novel IBFD, which have spurred a lot of activity in the last few years \cite{Bala2015}, \cite{Hanzo2016}. 

\section{Performance Assessment of 4D Constellations}
In the following, we describe the system model assumed for the performance assessment, as well as the chosen performance metric. Finally, we provide some details about the investigated constellations.
\subsection{System Model}
\begin{figure}
  \centering
    \includegraphics[width=1\linewidth]{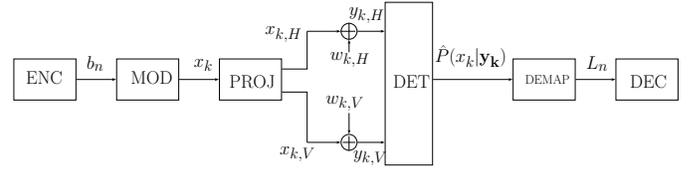}
	\vspace{-0.5 cm}
    \caption{System model for the performance assessment of 4D constellations.}
    \label{fig:block_4D}
\end{figure}

The considered system model is depicted in Fig.\,\ref{fig:block_4D}. The information symbols $\{x_{k}\}$ belong to a 4D constellation $\chi$ having $M=2^{m}$ symbols, which are associated to the bits $\{b_{n}\}$ through the labeling $\mu\!\!:\!\!\chi\!\rightarrow\!\{0,1\}^{m}$. We denote by $\mu^{i}(x_{k})$ the value of the $i$-th bit of the label mapped to symbol $x_{k}$. Since the transmission on the physical channel is separate for each polarization, the selected 4D symbol $x_{k}$ has to be projected onto two orthogonal 2D planes. This operation generates two 2D symbols, $x_{k,RH}$ and $x_{k,LH}$, which are to be transmitted over the right-hand (RH) and left-hand (LH) side circular polarizations, respectively. In the following, we assume the information symbols $\{x_{k}\}$ to be independent and uniformly distributed random variables. Depending on the chosen 4D constellation $\chi$, the projected symbols $x_{k,RH}$ and $x_{k,LH}$ may be correlated and have a non-uniform probability distribution \cite{KhKa93}. Indeed, if the 4D constellation is not obtained as the Cartesian product of two 2D constellations, then the projection onto a 2D plane may induce a shaping in the projected 2D constellations. 
We denote by $x_{k}$ the 4D transmitted symbol at time $k$, and by $\mathbf{x}_{k}=[x_{k,RH},x_{k,LH}]^{T}$ the corresponding vector containing the 2D projected symbols transmitted on the two polarizations at time $k$. Hence, the received symbol reads $\mathbf{y}_{k}=[y_{k,RH},y_{k,LH}]^{T}=\mathbf{x}_{k}+\mathbf{w}_{k}$ where $\mathbf{w}_{k}=[w_{k,RH},w_{k,LH}]^{T}$ denotes the samples of the AWGN process introduced by the channel. On each polarization, we assume the additive noise component $w_{k,c}$ to be a circularly-symmetric complex Gaussian random variable with mean zero and variance $\sigma^{2}$ per component, where $c$ identifies the two polarizations. 
Since the possible correlation between $x_{k,RH}$ and $x_{k,LH}$ may cause a performance loss if the detection is performed separately on each polarization, only joint detection will be considered. The chosen detection strategy is the soft maximum likelihood, providing as output of the detector, at every time $k$, the set of $M$ a posteriori probabilities $\{P(x_{k}|\mathbf{y}_{k})\}$ \cite{PrSa08}.
 
\subsection{Pragmatic Achievable Information Rate}
In order to assess the constellation performance, the chosen metric is the PAIR \cite{CaTaBi99}, defined as 
\begin{equation}
I_{p}(\chi,\mu)=\frac{1}{M}\sum_{i=1}^{m}\sum_{x\in\chi}E_{w}\left\{ \log\frac{P(y|\mu^{i}(x))}{P(y)}\right\}
\label{eq:pragm2}
\end{equation}
where the expectation is taken with respect to the AWGN distribution. Since a closed-form expression for the expectation in \eqref{eq:pragm2} is not available, numerical methods are usually employed for its evaluation.

\subsection{4D Constellation Designs}
In the following, the two classes of 4D constellations are described and their main features outlined.
\subsubsection{Cartesian Constellations}
The simplest way to obtain a $M$-ary 4D constellation is by taking the Cartesian product of two 2D constellations (e.g., $\sqrt{M}$-QAM), called
constituent constellations \cite{KhKa93}. Since $\sqrt{M}\!\!\times\!\!\sqrt{M}$-QAM are equivalent to MIMO with spatial multiplexing, they will be used in the following as performance benchmark. 
It is worth noting that the projection is not necessary in DP systems using Cartesian-based 4D constellations. Indeed, since the projection can be viewed as the inverse of the Cartesian product, the projected constellations coincide with the constituent constellations. 

For $\sqrt{M}\!\!\times\!\!\sqrt{M}$-QAM, the mapping has been obtained by applying the Gray mapping separately on the two 2D constituent constellations.

\subsubsection{Non-Cartesian Constellations}
The remaining types of constellations considered in this paper, i.e., LAM and EPPM, belong to this class. Since the construction of EPPM is detailed in \cite{YoWeYaBaKo13}, it is not reported here. The adopted mapping has been obtained numerically by using an instance of the generic algorithm aiming at maximizing the PAIR. 

The densest lattice in 4D is the centered cubic lattice $D_4$, the so-called checkerboard \cite{CoSl99}. Being the densest lattice in 4D, it provides the best coding gain by maximizing the minimum
Euclidean distance between the constellation symbols \cite{CoSl99}. In order to get a $M$-LAM, only $M$ points of the lattice are selected. The selection is done by choosing the points closest to the origin, i.e., the ones with the lowest energy. This is equivalent to perform a cut of the lattice with a centered sphere and selecting the lattice points inside the sphere. Such a spherical cut induces a shaping gain in the projected 2D constellations. If the spherical cut selects more than $M$ points, it means that many $M$-ary constellations with different average energies exist. Then, an instance of the genetic algorithm is applied over the set of possible constellations with the lowest energy in order to find the one with the highest AIR.   

For $M$-LAM, because of the high number of neighboring points \cite{CoSl99}, no Gray mapping exists. Moreover, for some values of $M$ it is possible to construct a quasi-Gray mapping (QG) of order 2 or 3 (where QG($n$) denotes a mapping where the labels of neighboring symbols differ for at most $n$ bits), but in general the corresponding performance is worse than that obtained with a numerically optimized mapping.  

\section{Adaptive MIMO Scheme, Modulation Order and Coding Rate in Dual Polarized in Satellite Communications}
\subsection{Physical Layer Abstraction}
\label{sect:dualpol}
The goal of PLA is to obtain the instantaneous error rate in order to estimate the instantaneous of capacity as a function of the radio channel coefficients. Hence, it is possible to run the simulations where the channel fading may not be correlated in time and therefore speed up the time of simulation. The model takes the modulation scheme, the coding rate, polarization scheme and many other parameters to adjust the bit loading depending on the magnitudes of the radio channel. PLA also offers the chance to study and analyse the impact of the feedback carried by UE.

Since the conveyed symbols are convoluted by the channel, each symbol experiments a different channel fading and therefore the Signal to Interference plus Noise Ratio (SINR) is different between the symbols in the same block. Thus, a metric of effective SINR is needed. This metric maps the equivalent SINR of the transmitted block to the error rate and it is called effective SINR mapping (ESM). Hence, the ESM is defined as a function to obtain the error rate from a single value that represents the effective SINR.
From~\cite{MoisioOborina2006}, the effective SINR is mathematically defined as
\begin{equation}
\bar{\gamma}=\beta_1\phi^{-1}\left(\frac{1}{N}\sum_n^N\phi\left(\frac{\gamma_n}{\beta_2}\right)\right)
\end{equation}
where $\boldsymbol{\gamma}$ is the $N$-length vector of the SINR of each symbol and $\beta_1$, $\beta_2$ are parameters to adjust the accuracy of the approximation. The function $\phi(.)$ defines the approach of ESM. 

In some cases, the representation of the error curves does not contain an analytical expression or becomes too complex. Thus, different approaches are proposed in the literature. In this paper, we use Mutual Information Effective SINR Mapping (MIESM) since it takes the function of the capacity of the link and estimates the equivalent SINR. It is expressed as
\begin{equation}
\phi(x)=\Expect_{XY}\left\{\log_2\frac{P\left(Y|X,x\right)}{\sum_{X'}P\left(X'\right)P\left(Y|X',x\right)}\right\}
\label{eq:miesm}
\end{equation}
where $X$ is the transmitted symbol, $Y$ is the received symbol and $\Expect\left\{.\right\}$ is the expected value. Assuming that a symbol is transmitted with a $M$-ary constellation,~\eqref{eq:miesm} can be expressed as
\begin{equation}
\phi(y)=\log_2M-\frac{1}{M}\sum_{x\in X}\Expect_w\left\{\log_2\sum_{x'\in X}e^{-\frac{|x-x'+w|^2-|w|^2}{\sigma^2}}\right\}
\label{eq:miesm2}
\end{equation}
where $X$ is the set of the constellation and $w\sim\mathcal{CN}(0,\sigma^2)$ and $\sigma^2=1/\gamma$.
This expression can be computed offline via Montecarlo simulations generating different realizations of the random variable. Nevertheless, in~\cite{IEEE2008} different results are exposed by QPSK, $16$QAM and $64$QAM, for a range of $-20:0.5:27$ dB of SINR. Although there is not a closed expression, it is possible to compute this expression for different values and store the results in a lookup table (LUT) to find the values of $\phi^{-1}(x)$~\cite{Rico-AlvarinoArnauMosquera2014}.

\subsection{Physical Layer Abstraction and MIMO}
In the previous section we described the PLA for the Single-Input Single-Output (SISO) and Single-Input Multiple-Output (SIMO) scenarios. In the case, the performance of the previous abstraction depends on the implementation of the receiver. In~\cite{IEEE2008} propose two approaches depending on the receiver:
\begin{itemize}
\item Linear MIMO Receivers. The use of linear receivers allows low computational complexity implementations and offers the chance to suppress or mitigate the cross interference of the inputs. Thus, without loss of generality, the receiver can decouple both polarizations into two separate streams. Hence, the mapping is performed using the SISO/SIMO approaches.
\item Maximum Likelihood (ML) Receivers. In this approach,~\eqref{eq:miesm2} is rewritten as a function of the probability of log-likelihood ratio (LLR). However, this approach requires much more computational complexity and requires additional LUTs, which enlarges the required memory.
\end{itemize}
For a given $n$th symbol, the system model of $t$ inputs and $r$ outputs MIMO scenario is described as
\begin{align}
\mathbf{y}_n&=\mathbf{H}_n\mathbf{x}_n+\mathbf{w}_n\\
\mathbf{H}&=\left(\begin{smallmatrix}\mathbf{h}_1&\mathbf{h}_2 \end{smallmatrix}\right)=\left(\begin{smallmatrix}h_{11}&h_{12}\\h_{21}&h_{22}\end{smallmatrix}\right)
\end{align}
where $\mathbf{y}\in\mathbf{C}^r$ is the received vector, $\mathbf{H}\in\mathbf{C}^{r\times t}$ is the random channel matrix, $x\in\mathbf{C}^t$ is the transmitted vector and $w\sim\mathcal{CN}\left(0,\sigma_w^2\mathbf{I}_r\right)$ is the additive white Gaussian noise (AWGN). In order to guarantee a feasible implementation, we use the linear MIMO receivers approach. 

MIESM for linear receivers is described by the SINR expression depending on the MIMO scheme:
\begin{itemize}
\item SISO: a single polarization is used. Thus, the system model is expressed as $y_n=h_n x_n+w_n$ and therefore $\gamma_n=|h_n|^2/\sigma_w^2$.
\item Orthogonal Polarization Time Block Codes (OPTBC): adaptation of the Orthogonal Space Time Block Codes, introduced in~\cite{Alamouti1998}, replacing the spatial component by the polarization component. The Since the OPTBC scheme exploits the full diversity of the channel~\cite{KaltenbergerLatifKnopp2013}, the SINR can be expressed as $\boldsymbol{\gamma}_n=\|\mathbf{H}_n\|^2/\sigma_w^2$, where $\|\mathbf{H}_n\|^2$ is the Frobenius norm.
\item Polarization Multiplexing (PM): each polarization conveys a symbol and thus, two symbols are transmitted in each channel access. Assuming that the receiver is able to cancel the interference between both streams, we obtain two equivalent SINR for each symbol of each polarization~\cite{KaltenbergerLatifKnopp2013,LatifKaltenbergerNikaeinEtAl2013}. Hence, it is equivalent to the previous section (the SISO case) but with $2N$ symbols rather than $N$. Therefore, the equivalent SINR of the $m$th polarization using the Zero Forcer (ZF) receiver is expressed as $\boldsymbol{\gamma}_{n,m}=\mathbf{h}_{n,m}^H\mathbf{h}_{n,m}/\sigma_w^2$
\item Polarized Modulation (PM): a single symbol is transmitted using a single polarization but the index of the used polarization is also a place for conveying bits. In the case where two polarizations are used, PMod conveys $M+1$ bits ($M$ bits of the symbol and an additional bit of the polarization state index)~\cite{HenarejosPerez-Neira2015a,HenarejosPerez-Neira2015}. Since a single symbol is transmitted, the received SINR is equivalent to SISO expression.
\end{itemize}

\subsection{Physical Layer Abstraction and BGAN}
After the introduction of the PLA for MIMO schemes, we aim to implement it to the BGAN standard. This standard describes different modulation and coding schemes (MODCOD), with different modulation schemes and different coding rates called bearers. Each bearer defines a MODCOD, which has different bit rate. From this standard, we get that the length of the block, $N$, can be $640$, $1098$ or $941$; and the constellation size $M$ can be $2$, $4$, $5$ or $6$. It is important to remark that since each MIMO scheme produces a different SINR, for the same channel realization each MIMO scheme will produce a different error curve.
Said that, we can formulate the objective problem as
\begin{align}
&\max_{u_{m,d,c}}\sum_{m\in\mathcal{M}}\sum_{d\in\mathcal{D}}\sum_{c\in\mathcal{C}}u_{m,d,c}r_{m,d,c}\left(\bar{\gamma}\right)\\
s.t.&PER\left(\bar{\gamma}\right)\leq 10^{-3}\\
&\sum_{m\in\mathcal{M}}\sum_{d\in\mathcal{D}}\sum_{c\in\mathcal{C}}u_{m,d,c}=1
\end{align}
where $\mathcal{M}$ is the set of MIMO modes, $\mathcal{D}$ is the set of modulation orders and $\mathcal{C}$ is the set of available coding rates, $r_{m,d,c}\left(\bar{\gamma}\right)$ is the achievable rate given the effective SNR $\bar{\gamma}$ and the tuple $m,d,c$. In this paper, we constraint the PER less or equal to $10^{-3}$.

\section{Numerical Results}
In this section we present the numerical results obtained in the different scenarios.
\subsection{4D Constellations in Fixed Scenarios}
\begin{figure}
  \centering
    \includegraphics[width=1\linewidth]{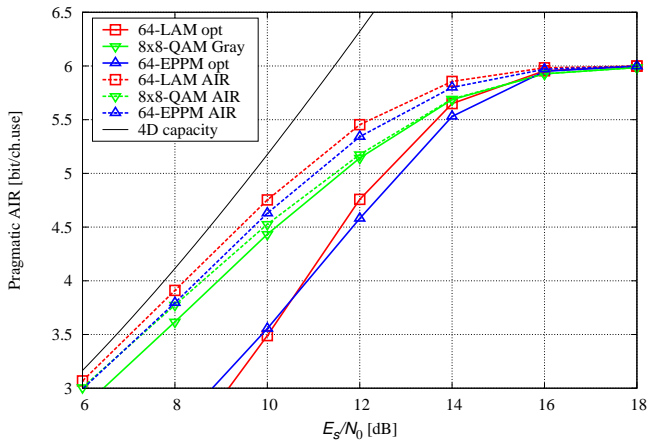}
	\vspace{-0.7 cm}
    \caption{AIR and PAIR for 64-ary 4D constellations.}
    \label{fig:air_pair}
\end{figure}

\begin{figure}
  \centering
    \includegraphics[width=1\linewidth]{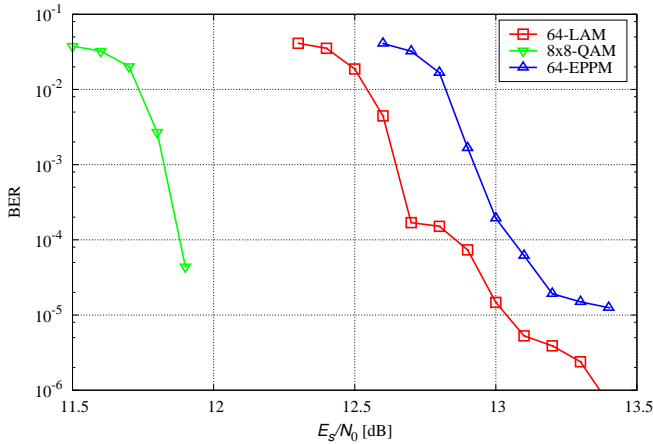}
	\vspace{-0.7 cm}
    \caption{BER for 64-ary 4D constellations.}
    \label{fig:4D_ber}
\end{figure}

For lack of space, we report the results obtained for $M=64$ only. Fig.\,\ref{fig:air_pair} shows the performance of $8\!\times\!8$-QAM, 64-LAM, and 64-EPPM in terms of AIR and PAIR. Concerning the AIR, 64-LAM and 64-EPPM outperform $8\!\times\!8$-QAM, showing gains stemming from the shaping of the corresponding 2D projected constellations caused by their non-Cartesian nature. However, when the mapping is applied (Gray for $8\!\times\!8$-QAM, numerically optimized for 64-LAM and 64-EPPM), non-Cartesian constellations show an impressive loss with respect to their AIRs (around 2 dB at 5 bits/ch.use). These results are validated by the bit error rate (BER) curves shown in Fig.\,\ref{fig:4D_ber}, where a LDPC code with rate $5/6$ has been used. The error floors in the curves relative to 64-LAM and 64-EPPM testify that the chosen code is not suitable for these non-Cartesian constellations. Better results can be obtained by using numerically optimized constellations over the AWGN channel \cite{KaMaSh15}, or by considering fading channels, where the robustness stemming from the correlation between the polarizations introduced by the shaping can be exploited \cite{MaKaSh16}.

\begin{figure}
  \centering
    \includegraphics[width=1\linewidth]{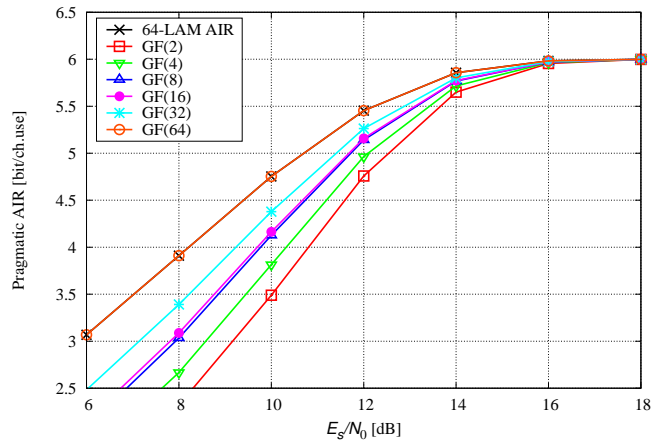}
	\vspace{-0.7 cm}    
    \caption{PAIR for 64-LAM with non-binary mappings.}
    \label{fig:lam_GF}
\end{figure}

Since the main responsible for the losses for LAM and EPPM is the binary mapping, we have tested non-binary mappings for 64-LAM. In particular, we have numerically optimized non-binary mappings over GF($2^n$) with $n=2,\dots,6$, where GF($q$) denotes the Galois field of order $q$. For each value of $n$, the mapping has been optimized by using the genetic algorithm to maximize the PAIR. Since $n>1$, in non-binary mappings each label is formed by $\ceil*{\log_{2^n}{M}}$ digits belonging to GF($2^n$), where $\ceil*{x}$ denotes the closest integer higher than $x$. The results reported in Fig.\,\ref{fig:lam_GF} show that increasing the order of the Galois field is beneficial, making the PAIR curve getting closer to the AIR curve. Moreover, when $n=6$, PAIR and AIR coincide. This is an expected result since using $n=6$ means that there is no mapping at all and the code shall operate directly on the 64-ary symbols, making AIR and PAIR equal. 

\subsection{2D Constellations in Mobile Scenarios}
\label{sect:results}
To simulate a mobile scenario, we generate a maritime scenario, where the user terminal is placed at the center of the satellite spot and is moving to the edge with a constant speed. During the trip, the terminal is receiving the blocks of symbols from the satellite and feedbacks the MODCOD and the MIMO mode, to optimize the throughput with a maximum $PER\leq 10^{-3}$. To make the simulations more realistic, we assume that there is a delay of $500$ms, which is a typical value.

To guarantee a fair comparison, first we generate a time series channel snapshot described in~\cite{SellathuraiGuinandLodge2006}, corresponding to $300$km trip. Later, we use this snapshot to run the different simulations for the different configurations and different user terminals.

Table~\ref{tab:simpara} describes the main parameters used in the system simulation.
\begin{table}[!ht]
\centering
\caption{Scenario Parameters}
\begin{tabular}{|r|l|}
\hline
Carrier	& $1.59$ GHz\\\hline
Beam Diameter &	$300$ km\\\hline
Noise &	$-204$ dBW/Hz\\\hline
Bandwidth & $32$ KHz\\\hline
TX Power & $4$ dBW\\\hline
Symbols per Block ($N$) & $640$\\\hline
Block Length & $20$ ms\\\hline
Channel Profile	& Maritime\\\hline
Speed of Terminal & $50$ km/h\\\hline
G/T	& $-13.5$ dB/K\\\hline
Feedback Delay & $500$ ms\\\hline
PLA Scheme & MIESM for MMSE\\\hline
\end{tabular}
\label{tab:simpara}
\end{table}

\begin{figure}[!ht]
  \centering
    \includegraphics[width=1\linewidth]{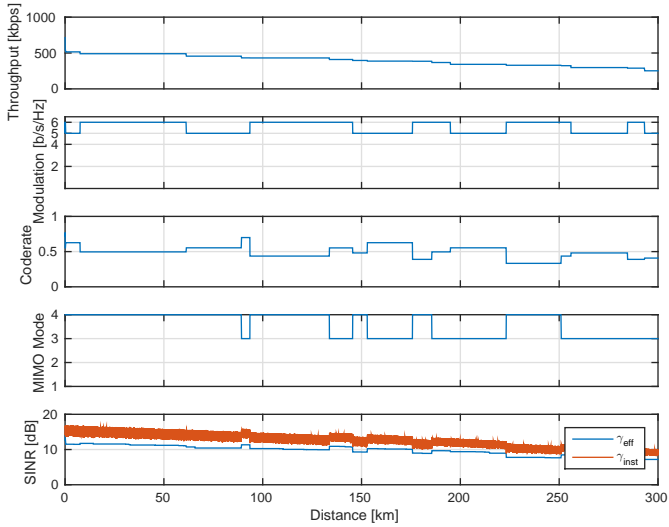}
    \caption{MODCOD and MIMO adaptation and delayed feedback of $500$ms.}
    \label{fig:res_all}
\end{figure}

Fig.~\ref{fig:res_all} summarizes the performance of the proposed adaptive techniques. In this figure, the adaptation between the modulation, coderate and also the MIMO mode is clear. For instance, during the major part of the trip, the transmitter uses the VBLAST scheme and, near the edge of the beam, uses the OPTBC scheme. 

\begin{figure}[!ht]
  \centering
    \includegraphics[width=1\linewidth]{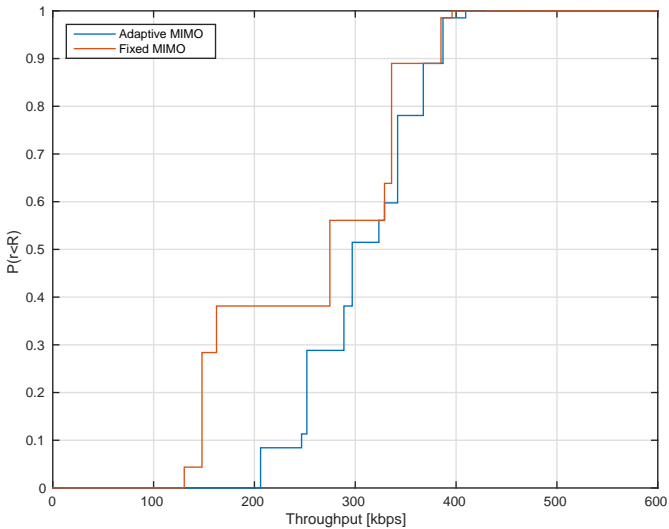}
    \caption{CDF of fixed MIMO and adaptive MIMO with delayed feedback of $500$ms.}
    \label{fig:res_cdf}
\end{figure}

Fig.~\ref{fig:res_cdf} compares the adaptive MIMO framework with the fixed V-BLAST scenario, i.e., where V-BLAST is always performed and the adaptation is done only through the MODCOD. In this case, the cummulative density function of the throughput is depicted.

\section{Full Duplex Communications}
This section  describes the operation of radio transceivers which operate in full-duplex mode and suffer from imperfect isolation between transmitter and receiver. The main source of degradation in IBFD transceivers is the coupled signal which the receiver picks from the transmitter. This signal, known as self-interference (SI),  can overwhelm the received signal from the other end, disrupting the communication if proper attenuation is not enforced. To a lesser extent, this problem arises also in FDD transceivers when the separation between transmission (TX) and reception (RX) frequency bands is not sufficient to guarantee the blocking of the coupling by passive filters.

\begin{figure}[!ht]
  \centering
    \includegraphics[width=1\linewidth]{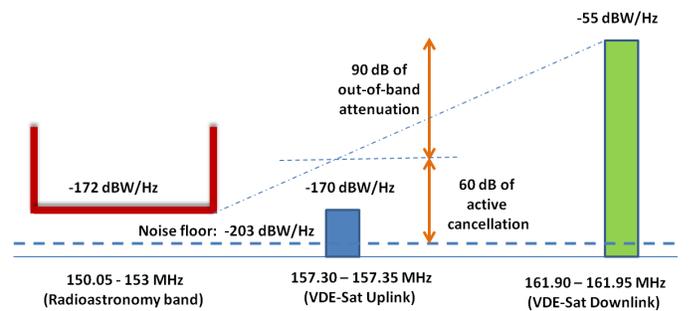}
    \caption{Frequency allocation in VDE-Sat. Self-interference attenuation needs amount to 150 dB.}
    \label{fig:freqalloc}
\end{figure}

The specific system which motivated this study is the spatial segment of VDES. The channelization in VDES is such that $50$ KHz are exclusively  assigned to the VDE-SAT downlink (channels $2026$ and $2086$), with  another  $50$ KHz (channels $1026$ and $1086$)  of exclusive use for the VDE-SAT uplink. The main source of interference in channels $1026$ and $1086$ is expected to come from the simultaneous transmission in the VDE-SAT downlink band. Even though terminals at ships will be half-duplex, without simultaneous transmission and reception, the standard does not preclude this option  in the satellite. Some additional complexity is expected in order to cope with the resultant interference, although the potential benefit is the increase in the spectral efficiency with respect to an alternating use of transmission and reception bands. The exclusive $50$ KHz bands for  the space-Earth link are reflected in Fig.~\ref{fig:freqalloc}, together with the radioastronomy emission mask  and the corresponding power spectral densities. Values  are taken from \cite{IALA}. The satellite receiver noise temperature is $25.7$ dBK without external interference, which gives $-203$ dBW/Hz. The received power is above that in the figure for ship elevation angles between 0 and 65 degrees. A simple extrapolation of the radioastronomy mask shows that the admissible emission in the VDE-SAT uplink RX band would be way too high with respect to the received signal magnitude. This out-of-band leakage coming from the transmitter reduces the sensitivity of the receiver. It also increases the linearity requirement of the RX front-end due to its significantly high power levels in comparison to the received signals. Furthermore, in the case that the TX  leakage is large, the LNA and mixer can then be forced into saturation resulting in desensitization of the RX frond-end. Conventionally, the out-of-band TX leakage is suppressed by placing selective band pass filters at the receiver. Active cancellation can relax filter specifications or can provide additional cancellation. In the case under consideration, passive RX filtering would not be enough if the out-of-band emissions are uniquely constrained by the radioastronomy service. An initial first-order rough approximation shows that the out-of-band noise should be 90 dB below the TX signal, and still 60 dB of attenuation would be additionally required in order to keep it at the level of the noise floor. This attenuation should be pursued by active  means as discussed next.

\subsection{Self-Interference Cancellation for FDD}
\begin{figure}[!ht]
  \centering
    \includegraphics[width=1\linewidth]{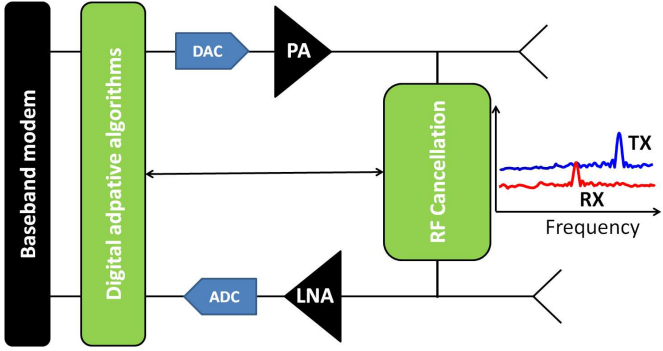}
    \caption{Block diagram of a full duplex transceiver including both analog and digital active cancellation units.}
    \label{fig:mixedcancellation}
\end{figure}

\begin{figure}[!ht]
  \centering
    \includegraphics[width=0.7\linewidth]{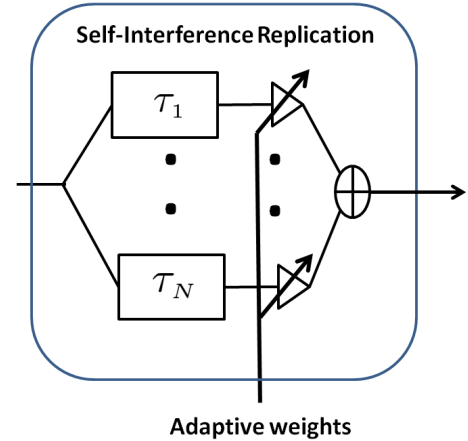}
    \caption{Fixed delays and variable attenuators to replicate the self-interfering signal.}
    \label{fig:delays}
\end{figure}

Fig.~\ref{fig:mixedcancellation} shows the operating principle of an FDD transceiver with active interference cancellation. Although two antennas are assumed, the  considerations to follow apply for a single antenna payload with a duplexer partially decoupling the RX from the TX. By tapping the reference signal from the power amplifier (PA) output, all non-idealities of the transmit chain are included. Automatic Gain Controls (AGC), not shown in the figure, are needed also for the management of the interference. The AGC performs as an attenuator when the SI is very high, during the initial adjustment of the adaptive parameters. Otherwise, the saturation of the front-ends would prevent convergence. 
The operating principle is the same for both analog and digital cancellation schemes. The estimation of the received SI waveform is needed to subtract its contribution from the received signal. The replication of the SI signal follows the well-known principles of echo cancellation, although the  implementation with RF waveforms presents some issues of its own. Fig.~\ref{fig:delays} shows how delay lines and weights are put together to form an analog filter. As opposed to a conventional digital filter, the number of analog delays or taps is constrained by their implementation complexity. Thus, the cancellation performance is limited by the degrees of freedom in terms of delays and their associated resolution, since high resolution programmable delays are very complex. A small number of high resolution delays makes sense if they can be adjusted. Otherwise, solutions with a large number of low-resolution delay lines are handier \cite{Bharadia2013}.  Even further, other solutions including the use of programmable bandpass filters could be considered \cite{Zhou2015}.

\begin{figure}[!ht]
  \centering
    \includegraphics[width=1\linewidth]{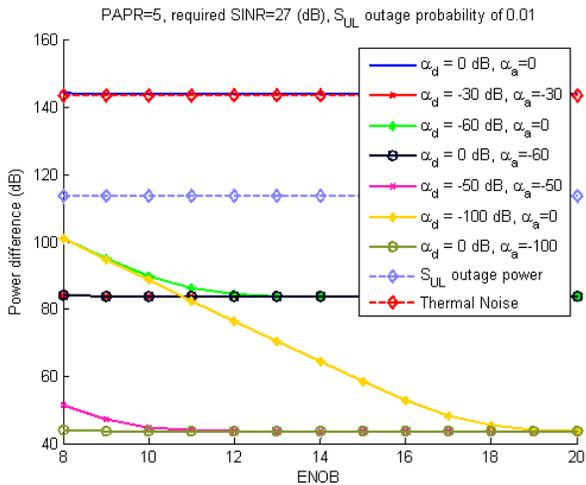}
    \caption{Self-interference levels with respect to input interference power.}
    \label{fig:mixedperformance}
\end{figure}

SI cancellation is expected to come from the complementary analog and digital cancellation efforts. The analog reduction of the SI needs to guarantee that the resultant dynamic range can be handled by the reception chain, including the ADC. In this regard, it is highly important to characterize the combined performance of this mixed architecture. If $\sigma_d^2,\sigma_n^2, \sigma_s^2, \sigma_e^2$ denote the desired signal power, the noise power, the SI signal power and the quantification noise power, respectively, and $\alpha_a$ and $\alpha_d$ denote the analog and digital cancellation factors, respectively, then the $\mbox{SINR}$ after cancellation can be written as \cite{Schniter2014} 
\begin{equation}
\mbox{SINR} = \frac{\sigma_d^2}{\sigma_s^2\cdot \alpha_a \cdot \alpha_d + \sigma_e^2 + \sigma_n^2}
\end{equation}
with
\begin{equation}
\sigma_e^2 = \mbox{PAPR} \cdot  (\sigma_s^2\cdot \alpha_a + \sigma_d^2 + \sigma_n^2)\cdot 10^{\frac{-6.02 \cdot \mbox{\tiny ENoB}}{10}}
\end{equation}
where $\mbox{PAPR}$ is the Peak-to-Average Power Ratio, and $\mbox{ENoB}$ is the effective number of bits of the ADC. These expressions are used to plot in Fig.~\ref{fig:mixedperformance} the required passive attenuation to achieve an $\mbox{SINR}=27$ dB, corresponding to the magnitudes shown in the spectral allocation in Fig.~\ref{fig:freqalloc}. This passive attenuation will come from propagation losses between transmit and received antenna, and from the relative level between the out-of-band noise and the in-band transmit power. Thus, if a cancellation of 60 dB is achieved ($30$ dB + $30$ dB), the leaked SI power in the receiver band should be almost $90$ dB lower than the transmit power in order to avoid any degradation. This 90 dB should come from the transmit chain selectivity and coupling losses. Thus, passive filters and propagation techniques will determine the self-Interference level at RF frond-end and how effective the active cancellation needs  to be.

In order to grasp the order of magnitude of the cancellation levels that we can expect, let us consider first a simple non-frequency selective channel, with a delay that we assume that can be perfectly matched by the cancellation circuit. That is, the only error will come from the estimated weight of the coupled signal. Thus, if $s(t)$ denotes the interfering signal, the residual echo is given by 
\begin{equation}
e(t) = s(t) - (1-\epsilon)s(t)
\end{equation}
with $\epsilon$ the relative estimation error. Echo cancellation performance is measured as 
\begin{equation}
\mbox{Cancellation [dB]} = 10 \log_{10} \frac{\mathbb{E}[|s^2(t)|]}{\mathbb{E}[|e^2(t)|]},
\end{equation}
which for a magnitude error $\epsilon$ is equal to $-20\log_{10}(|\epsilon|)$. The cancellation performance degrades significantly even with low errors in the estimation of the coupling magnitude; thus, for example, a $1\%$ relative error bounds  the cancellation performance to 40 dB. Additional degradation will come from the mismatched delay.  This cancellation level can be shown to be given by 
\begin{equation}
\mbox{Cancellation [dB]} = -20 \log_{10} (1-\mbox{sinc}(B\tau))
\end{equation}
where $B$ and $\tau$ denote the signal bandwidth and time error, respectively. More generally, the coupling attenuation will depend on the accuracy of the estimation of the transfer function, for which a good modelling of the coupling path is essential. In addition, practical implementation requires the automatic tuning of the involved parameters, assuming that the required cancellation level can be achieved by a proper selection of delay lines and time resolution. 

The adaptive algorithms which are used to adapt the weights are usually taken  from the well-known LMS family of algorithms, keeping in mind that filtering takes place with RF waveforms and adaptation in digital baseband. 

On the other side, digital cancellation should be able to target cancellation levels about $30$ dB. In fact, this is the reference which can be achieved in a well-known concept in SatCom known commercially as Paired-Carrier Multiple Access or DoubleTalk Carrier-in-Carrier, and conceived to exchange information between two ends through a satellite, by making use of the same carrier in the uplink. The cancellation of the local signal requires a digital cancellation step, although no analog cancellation module is needed since the uplink and downlink carriers are different \cite{Treichler15}.

\section{Conclusions}
\label{sect:conclusions}
In this paper the performance of different 4D constellations has been assessed. We showed that non-Cartesian designs provide gains only in uncoded systems, since in coded system binary mappings cause relevant losses. Such losses can be highly reduced by resorting to non-binary mappings which, however, require non-binary codes, whose complexity (especially regarding the decoder) is still formidable and makes them more suitable for theoretical investigation rather than practical implementation.

Furthermore, we also analysed the impact of the use of polarization in mobile scenarios. We introduced the concept of PLA applied to the MIMO case. Benchmarked to BGAN standard, we proposed the use of PLA to model the adaptive downlink based on the effective SNR at the terminal. We showed the gain and benefits of adapting the modulation order, coding rate and MIMO scheme jointly, specially in low SNR regimes. And finally, we analysed the viability of this technique benchmarked to the BGAN standard.

With respect to interference cancellation in full-duplex communications, no  stand-alone analog or digital technique is usually capable of achieving a cancellation level high enough to satisfy the demanded performance. In this regard, it is required to effectively balance the roles of the analog and digital cancellation, together with the passive suppression in the form of filtering and antenna decoupling. Finally, hardware imperfections, not mentioned here, can impose some performance limitations on the SI cancellation. 

\section*{Acknowledgement}
Part of this work has been performed in the framework of the SatNEx 4, which is funded by the European Spatial Agency. This work has received funding from the Spanish Ministry of Economy and Competitiveness (Ministerio de Economia y Competitividad) under project TEC2014-59255-C3-1-R and from the Catalan Government (2014SGR1567).

\bibliographystyle{IEEEtran}
\bibliography{biblio}

\end{document}